\documentclass[aps,prb,twocolumn,superscriptaddress,showpacs]{revtex4}

\usepackage{graphicx}
\usepackage{amsfonts}
\usepackage{amssymb}
\usepackage[]{amsmath}
\usepackage[]{latexsym}
\usepackage[]{float}
\usepackage{bm}

\begin{document}


\title{Anomalous 
criticality in the quantum Hall transition at $n=0$ Landau level of graphene\\
with chiral-symmetric disorders}

\author{Tohru Kawarabayashi}
\affiliation{Department of Physics, Toho University,
Funabashi, 274-8510 Japan}

\author{Takahiro Morimoto}
\affiliation{Department of Physics, University of Tokyo, Hongo, 
Tokyo 113-0033 Japan }

\author{Yasuhiro Hatsugai}
\affiliation{Institute of Physics, University of Tsukuba, Tsukuba, 305-8571 Japan}

\author{Hideo Aoki}
\affiliation{Department of Physics, University of Tokyo, Hongo, 
Tokyo 113-0033 Japan }

\date{\today}

\begin{abstract}
We investigate numerically whether the chiral symmetry is 
the sole factor dominating the criticality 
of the quantum Hall  transitions in disordered graphene.  
When the disorder respects the 
chiral symmetry, the plateau-to-plateau transition at the $n=0$ Landau level is shown to become anomalous (i.e., step-function-like).   
Surprisingly, however, 
the anomaly is robust against the inclusion of the 
uniform next-nearest neighbor hopping, which  degrades the chiral symmetry of the lattice models. 
We have also shown that the ac (optical) Hall conductivity exhibits 
a robust plateau structure when the disorder respects the chiral symmetry. 
\end{abstract}


\maketitle

\section{Introduction}
Kicked off by the recent observation of the characteristic quantum Hall effect in graphene,\cite{Geim,Kim,ZhengAndo} the effect of disorder has 
attracted considerable attention.\cite{KA,SM,OGM,KHA}  
A key interest that is specific to the  electronic structure of graphene 
is the chiral (A-B sub-lattice) symmetry \cite{HFA,HChiral} and the scattering between two Dirac cones that reside at K and K' points in k-space.  
Accordingly, the nature of disorder (i.e., whether the disorder preserves 
the chiral symmetry or not) should be essential in discussing the electronic properties of disordered graphene.
A potential disorder breaks the chiral symmetry, whereas a disorder in bonds preserves the symmetry.  
The spatial range of disorder is another important factor, since 
the inter-valley scattering  occurs when the disorder is 
short-ranged while strongly suppressed when long-ranged.  

In our previous papers \cite{KHA}, we have considered the disorder in the 
hopping energies as 
a model for a ripple,\cite{Meyer,GLERSRLM} which is an intrinsic disorder in graphene.  We have 
clearly demonstrated that for such a chiral-symmetry preserving 
disorder, the criticality of the quantum Hall transition at the 
$n=0$ Landau level becomes anomalously sensitive to the spatial correlation of disorder:   
When the spatial range of the bond disorder exceeds a few lattice constant, no broadening of the  graphene-specific $n=0$ Landau level
occurs and the associated step-function like Hall transition exhibits an anomalous (almost fixed-point) criticality.
This sharply contrasts with the 
case of potential disorder which degrades the chiral symmetry, where the criticality is similar to the ordinary quantum Hall
transition.\cite{NRKMF}  

Now, when one describes the actual graphene with a tight-binding model, the effect of the next-nearest neighbor hopping in the honeycomb lattice is not 
negligible.\cite{Neto,Reich,Nakajima} 
Since the next-nearest neighbor hopping $t'$ 
connects sites in the same sub-lattice, the chiral symmetry of the original the honeycomb lattice is broken, 
so that one might naively think that the speciality about the 
$n=0$ Landau level associated with the chiral symmetry should be 
washed away.  It is to be noted, however, that  the space group 
(i.e., rotation and reflection symmetries as well as the translation 
symmetry) of the 
honeycomb lattice are preserved, and, indeed,  the two Dirac cones still exist at $K$ and $K'$ points at an energy shifted from zero.\cite{SW}
The existence of the doubled Dirac cones has a topological origin.  
For instance, they exist even in the presence of a third-neighbor hopping 
$t''$ in one
of the three directions (which is equivalent to the 
$\pi$ flux model) with $t'=0$ as shown in Ref. \onlinecite{HFA}, where the lattice 
continuously crosses over to the square lattice as $t''$ is 
increased.   In such a system, the Dirac cones are located at general points away from K and K' in the Brillouin zone.   
If we add $t'$ in this case, 
tilted Dirac cones appear at a non-zero energy. 
The tilted Dirac cones are interesting, since they have been 
observed in an organic system $\alpha$-(BEDT-TTF)$_2$I$_3$ under a high pressure.\cite{KKS,TSTNK}   

Given these variety of situations, we want to clarify 
in the present paper whether the anomalous criticality is preserved or washed out.  First, we shall clarify the question: 
Do we have to discriminate the presence or otherwise of 
the chiral symmetry in the underlying honeycomb lattice (before 
we turn on the disorder) and  the presence or otherwise of 
the chiral symmetry in the disorder?  
In order to study these, we adopt the honeycomb
lattice model rather than the effective Dirac model, since the 
chiral symmetry has to do with the AB sublattices.  
To examine the robustness 
of the criticality for the $n=0$ Landau level, we calculate the Hall conductivity 
in the honeycomb lattice as well as the $\pi$ flux model with the bond disorder to examine the influence of 
the next-nearest neighbor hopping.
We find  
that the anomalous criticality at the $n=0$ Landau level for the long-ranged bond disorder 
is robust against the inclusion of the 
uniform next-nearest neighbor hopping, namely the break down of the chiral 
symmetry of the lattice models, which yields, in general,  the energy shift as well as the tilt of the Dirac cones. 
The Hall conductivity for non-interacting electron systems is mathematically equivalent to a topological integer called a Chern number. 
\cite{TKNN,NTW,Hedge,AokiAndo}
Taking into account the recent progress in numerical evaluation of the Chern number, \cite{HC,FHS,HFS} we adopt 
a formulation based on the lattice gauge technique to evaluate the Hall conductivity in disordered finite systems. The formulation is 
based on the topological invariant used in the lattice gauge theory, which 
can also be understood as a two dimensional generalization
of the King-Smith-Vanderbilt formula 
for polarization.\cite{FHS,KSV}  The present formulation has been successfully applied  to the calculation of  the Hall conductivity 
not only for uniform systems but also for disordered systems.\cite{SMH}

Our second question is: How the criticality 
appears in the ac (optical) Hall conductivity, $\sigma_{xy}(\omega)$?  
The optical response of QHE system is now attracted growing interest.
For instance, the cyclotron emission from graphene QHE system has been
studied theoretically \cite{morimoto-CE}
as well as experimentally  with infrared transmission.\cite{sadowski06}
Specifically, Morimoto et al\cite{morimoto-opthall} have shown that 
the optical Hall conductivity $\sigma_{xy}(\omega)$ has 
a Hall plateau structure even in the ac ($\sim$ THz) regime, 
both in the ordinary two-dimensional electron gas (2DEG) 
and in graphene 
in the quantum Hall regime, 
although the plateau height is no longer quantized in ac.  
In graphene $\sigma_{xy}(\omega)$, which 
reflects the unusual Landau level structure, the plateau structure 
remains against a significant strength of disorder due to 
an effect of localization.  
For 2DEG the THz spectroscopy  has recently been performed, 
and a plateau structure of optical Hall conductivity has been detected in the THz region.\cite{ikebe-THz10}  
Hence it is interesting 
to investigate the  consequence of the chiral symmetry in the ac Hall effect in graphene.

\section{Chiral symmetry in lattice models}
Let us begin with the chiral symmetry.  
The symmetry is defined by the existence of a local operator $\gamma$ 
with  $\gamma^2=1$ 
that anti-commutes with the Hamiltonian $H$ as $\{ \gamma , H\}=0$. 
For bipartite lattices such as honeycomb and 
square, the lattice can be decomposed into $A$ and $B$ sub-lattices. 
If we take $\gamma = \prod_{i\in A} \exp({\rm i}\pi c_i^\dagger c_i)$, 
the fermion operator $c_i$ is transformed as 
$\gamma c_i \gamma = -c_i$ for $i \in A$ and $\gamma c_i \gamma = c_i$ for $i \in B$.
We can readily show that the lattice model that has only hopping terms $t_{ij}c_i^\dagger c_j$ connecting A and B sites 
has the chiral symmetry $\{\gamma ,H\}=0$, irrespective of the details of the hopping terms $t_{ij}$.

An important consequence of the chiral symmetry is that the energy levels appear  in pairs, $\{E, -E\}$, and hence 
the energy spectrum is symmetric around $E=0$. Assume that $\psi_E$ is an eigenstate with energy $E$ satisfying 
$H\psi_E = E\psi_E$.   Then the state $\gamma \psi_E$ is an eigenstate with energy $-E$, since 
$H(\gamma \psi_E) = -\gamma H\psi_E = -E(\gamma\psi_E)$. For $E=0$ 
the eigenstates $\psi_{E=0}$ 
and $\gamma\psi_{E=0}$ are degenerated.  We can 
rearrange the eigenstates into  $\psi_{\pm}= (1\pm \gamma) \psi_{E=0}$ to make them simultaneous eigenstates
of $\gamma$ with $\gamma \psi_{\pm} = \pm \psi_{\pm}$. This implies that the system has an additional symmetry in the zero-energy space.
From the definition of $\gamma$, it is clear that the eigenstate $\psi_{+(-)}$ has its amplitudes only on the $B(A)$ sub-lattice sites.  
The criticality of the Hall transition at $E=0$ can thus be sensitive to the presence or absence of the chiral symmetry.\cite{AZ}

\section{DC Hall conductivity}

\subsection{Hamiltonian}
\begin{figure}
\includegraphics[scale=0.4]{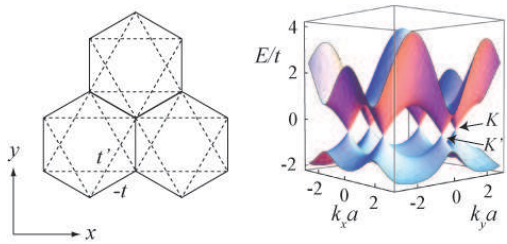}
\caption{
Left: Honeycomb lattice with 
nearest neighbor hopping $-t$ (solid lines) and 
next-nearest neighbor hopping $t'$ (dotted lines).   
Right: 
Energy dispersion for $\phi=0, t'/t=0.15$. Dirac cones are located at 
$K=(2\pi/3\sqrt{3}a, 2\pi/3a)$ and $K'=(4\pi/3\sqrt{3}a,0)$ with energy $-3t'$.\cite{Neto} 
\label{fig1}
}
\end{figure}

The tight-binding Hamiltonian for the honeycomb lattice with the next-nearest neighbor hopping is  given by 
$$
 H = \sum_{\langle i,j\rangle}^{\rm nn} t_{ij} e^{-2\pi{\rm i}\theta_{ij}} 
 c_i^{\dagger}c_j + \sum_{\langle i,j \rangle}^{\rm nnn} t' e^{-2\pi {\rm i}\theta'_{ij}}c_i^{\dagger}c_j + {\rm h.c.},
$$
where the summation is performed for nearest neighbor sites  in the first term and for next-nearest neighbor sites in the second term 
(Fig. \ref{fig1}). 
Here $t_{ij}$ and $t'$ are assumed to be real, 
while the Peierls phases, $\{ \theta_{ij} \}$ and $\{ \theta'_{ij}\}$, are determined such that
the sum of the phases around a loop is equal to the magnetic flux 
piercing the loop in units of the flux 
quantum $\phi_0=h/e$.  We specify the strength of the uniform magnetic field by the magnetic flux $\phi$ 
per hexagon in the honeycomb lattice. 
We introduce a bond randomness 
in the nearest-neighbor transfer energy as 
$ t_{ij} = -t + \delta t_{ij}$,
where $t$ and $\delta t_{ij}$ denote the uniform and the disordered 
components, respectively.  The disorder in 
$\delta t_{ij}$ is assumed to have a Gaussian distribution
$
 P(\delta t) = 
e^{-\delta t^2/2\sigma^2}/{\sqrt{2\pi \sigma^2}}
$  
with a variance $\sigma$.   
The spatial correlation length $\eta$ in the random components is specified
by requiring 
\begin{equation*}
 \langle \delta t_{ij} \delta t_{kl}
 \rangle = \langle \delta t^2 \rangle e^{-|\mbox{\boldmath $r$}_{ij} -
 \mbox{\boldmath $r$}_{kl}|^2/4\eta^2},
\end{equation*}
where $\mbox{\boldmath $r$}_{ij}$ represents the 
bond $t_{ij}$, and $\langle \rangle$ the ensemble 
average\cite{Kawa}. The length is hereafter measured in units of the distance $a$ between nearest-neighbor sites. 
It is to be noted that the Hamiltonian $H$ is chiral symmetric if $t'=0$ even in the 
presence of disorder in the nearest neighbor hopping $t_{ij}$, while the chiral symmetry is broken for $t' \neq 0$.
In actual graphene, the magnitude of $t'$ is estimated to be $0.2 \sim 0.02 t$.\cite{Reich}

\subsection{Hall conductivity as a Chern number}
In order to evaluate the Hall conductivity, let us consider a 
$L_x \times L_y$ system, and impose boundary conditions for the single-particle wave function $|\varphi \rangle$ as 
$
 T_{x(y)}(L_{x(y)})|\varphi  \rangle = e^{{\rm i}\phi_{x(y)}}|\varphi \rangle, 
$
where $T_{x(y)}(L_{x(y)})$ denotes the single-particle magnetic translation operator in the $x(y)$ direction.
The string gauge in the corresponding brick-layer lattice is adopted to treat this twisted boundary conditions. \cite{HIM} 
When the energy gap exists above the Fermi energy $E_F$,
the Hall conductivity $\sigma_{xy}$ can related to a topological integer $C_E$ called the Chern number  as 
\cite{HFA,TKNN,NTW,Hedge,AokiAndo,HIM}
$
  \sigma_{xy} =-\frac{e^2}{h}C_E 
$
with 
$$
 C_E = \frac{1}{2\pi{\rm i}} \int_{T^2} d^2\mbox{\boldmath $\phi$} (\partial_{\phi_x} \langle \psi | \partial_{\phi_y} \psi \rangle -
   \partial_{\phi_y}\langle \psi | \partial_{\phi_x} \psi \rangle) ,
$$
where $\psi$ denotes the wave function of the Fermi sea in the 
twisted boundary conditions specified by the phase $\mbox{\boldmath $\phi$}=(\phi_x,\phi_y), (0\leq \phi_{x(y)} \leq 2\pi)$ with the 
normalization $\langle \psi | \psi \rangle=1$
and $T^2$ is the two-dimensional parameter space for the phase $\mbox{\boldmath $\phi$}$.
The Chern number can be expressed, by introducing a $\delta \phi_x$ by $\delta \phi_y$ mesh for the parameter space $T^2$, as 
$
 C_E = \frac{1}{2\pi{\rm i}} \sum_{m,n} F_{m,n}
$
with 
$$
 F_{m,n} 
=\oint_{\partial S_{mn}} \mbox{\boldmath $A$} \cdot d\mbox{\boldmath $\phi$}, \quad  \mbox{\boldmath $A$}=\langle \psi | \nabla_{\phi} \psi \rangle.
$$
Here $S_{mn}$ stands for a $\delta \phi_x$ by $\delta \phi_y$ square of the mesh with points  $\mbox{\boldmath $\phi$} = (m\delta \phi_x, n\delta\phi_y),  ((m+1)\delta \phi_x, n\delta\phi_y), ((m+1)\delta \phi_x, (n+1)\delta\phi_y), (m\delta \phi_x, (n+1)\delta\phi_y)$. 
The surface integral on $S_{mn}$ is replaced by the line integral by  Stokes' theorem. 
It is useful to note that  this line integral $F_{m,n}$ can be written as  
$$
 F_{m,n}= \ln U_x (\mbox{\boldmath $\phi$})U_y 
 (\mbox{\boldmath $\phi$}+\mbox{\boldmath $\Delta \phi$}_x) U_x^{-1} (\mbox{\boldmath $\phi$}+\mbox{\boldmath $\Delta \phi$}_y)
 U_y^{-1} (\mbox{\boldmath $\phi$}),
$$
with 
$
 U_{\alpha} (\mbox{\boldmath $\phi$}) = \exp( \int_{\mbox{\boldmath $\phi$}}^{\mbox{\boldmath $\phi$}+\mbox{\boldmath $\Delta\phi$}_\alpha}
A_\alpha d \phi_\alpha ) \quad (\alpha = x,y),
$ 
where $\mbox{\boldmath $\Delta\phi$}_x=(\delta \phi_x,0)$, $\mbox{\boldmath $\Delta\phi$}_y=(0,\delta\phi_y)$ 
and $\mbox{\boldmath $\phi$} = (m\delta \phi_x, n\delta\phi_y)$.

Following Fukui et al.,\cite{FHS} we consider a link variable defined on 
the mesh as 
$$
 \tilde{U}_{\alpha} (\mbox{\boldmath $\phi$}) =  \frac{\langle \psi(\mbox{\boldmath $\phi$}) | \psi 
 (\mbox{\boldmath $\phi$}+\mbox{\boldmath $\Delta\phi$}_\alpha )\rangle}{|\langle \psi(\mbox{\boldmath $\phi$}) | \psi 
 (\mbox{\boldmath $\phi$}+\mbox{\boldmath $\Delta\phi$}_\alpha )\rangle|}, \quad (\alpha = x,y),
$$
and define the Chern number $\tilde{C}_E$ on the mesh as 
$
 \tilde{C}_E = \frac{1}{2\pi{\rm i}} \sum_{m,n} \tilde{F}_{m,n},
$
where
$$
 \tilde{F}_{m,n}= \ln \tilde{U}_x (\mbox{\boldmath $\phi$})\tilde{U}_y 
 (\mbox{\boldmath $\phi$}+\mbox{\boldmath $\Delta \phi$}_x) \tilde{U}_x^{-1} (\mbox{\boldmath $\phi$}+\mbox{\boldmath $\Delta \phi$}_y)
 \tilde{U}_y^{-1} (\mbox{\boldmath $\phi$})
$$
with $-\pi < \tilde{F}_{m,n}/{\rm i} \leq \pi$.
The Chern number $\tilde{C}_E$ defined in such a way on the mesh 
reduces to the Chern number $C_E$ defined 
on the continuum space in the 
limit $\delta\phi_{x(y)} \rightarrow 0$. It is to be noted that $\tilde{C}_E$ is gauge invariant, and 
takes  an integer value for an arbitrary mesh width $\delta\phi_{x(y)}$. 
We have then $\tilde{C}_E = C_E$ for a fine enough mesh. \cite{FHS,HFS} 

In the present case, the wave function of the Fermi sea $|\psi \rangle$ is a Slater determinant of  the single-particle wave function $|\varphi \rangle$, 
so that the link variable $\tilde{U}_\alpha(\mbox{\boldmath $\phi$})$ for an $M$ particle state is given by \cite{HC}
$
 \tilde{U}_\alpha (\mbox{\boldmath $\phi$}) = {\det V_\alpha (\mbox{\boldmath $\phi$})}/{|\det V_\alpha (\mbox{\boldmath $\phi$})|},
$
where $V_\alpha(\mbox{\boldmath $\phi$})$ is an $M \times M$ matrix. The elements of $V$ are defined by
$
 (V_\alpha(\mbox{\boldmath $\phi$}))_{ij} = \langle \varphi_i (\mbox{\boldmath $\phi$}) | \varphi_j(\mbox{\boldmath $\phi$}+
 \mbox{\boldmath $\Delta\phi$}_\alpha  ) \rangle, \  (1\leq i,j \leq M)),
$
with the $i$-th single-particle eigenstate $|\varphi_i (\mbox{\boldmath $\phi$}) \rangle$ satisfying $H|\varphi_i (\mbox{\boldmath $\phi$})\rangle = 
\varepsilon_i |\varphi_i (\mbox{\boldmath $\phi$}) \rangle$ ($\varepsilon_1 \leq \ldots \leq \varepsilon_k \leq \varepsilon_{k+1}, \ldots$) in the boundary condition specified by $\mbox{\boldmath $\phi$}$, where
the Fermi energy $E_F$ is assumed to lie between $\varepsilon_k$ and $\varepsilon_{k+1}$.
The energy gap at the Fermi energy is mostly guaranteed for finite  and random systems due to the level repulsion.
We thus adopt this formulation to evaluate the Chern numbers for each sample. By taking an average over samples in each energy bin,
we obtain ensemble-averaged Hall conductivity as a function of the Fermi energy.

\subsection{Anomalous criticality in the $n=0$ level}

\begin{figure}
\includegraphics[scale=0.3]{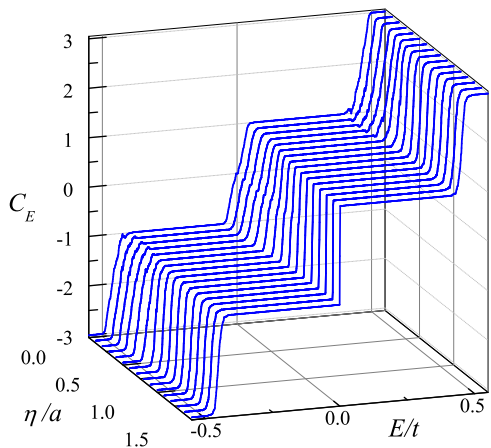}
\caption{(Color online) 
The Chern number $C_E$ (Hall conductivity in units of $e^2/h)$ as a function of the Fermi energy $E/t$ 
for various values of the spatial correlation length $\eta/a$ of the bond disorder, for  
$t'/t=0$, $\phi/\phi_0 = 1/50$, $\sigma/t = 0.115$ and $L_x/(\sqrt{3}a/2)=L_y/(3a/2) =20$. 
\label{fig2}
}
\end{figure}

The Chern number $C_E$ for the case of a bond-disordered honeycomb lattice with $t'=0$, where  the chiral symmetry 
is exactly preserved,  is shown in Fig.\ref{fig2}. 
We can clearly see that the plateau transition in the $n=0$ Landau level 
at $E=0$ becomes 
almost step-function like as a function of the Fermi energy 
as soon as the spatial correlation length, $\eta$, of the bond disorder exceeds the lattice constant  $a$, as previously shown in Ref. \onlinecite{KHA}.  
Remarkably, this behavior is fixed-point like in that it shows almost no system-size dependence.  
The anomalous Hall transition is accompanied by the absence of broadening of the $n=0$ Landau level.

\begin{figure}
\includegraphics[scale=0.3]{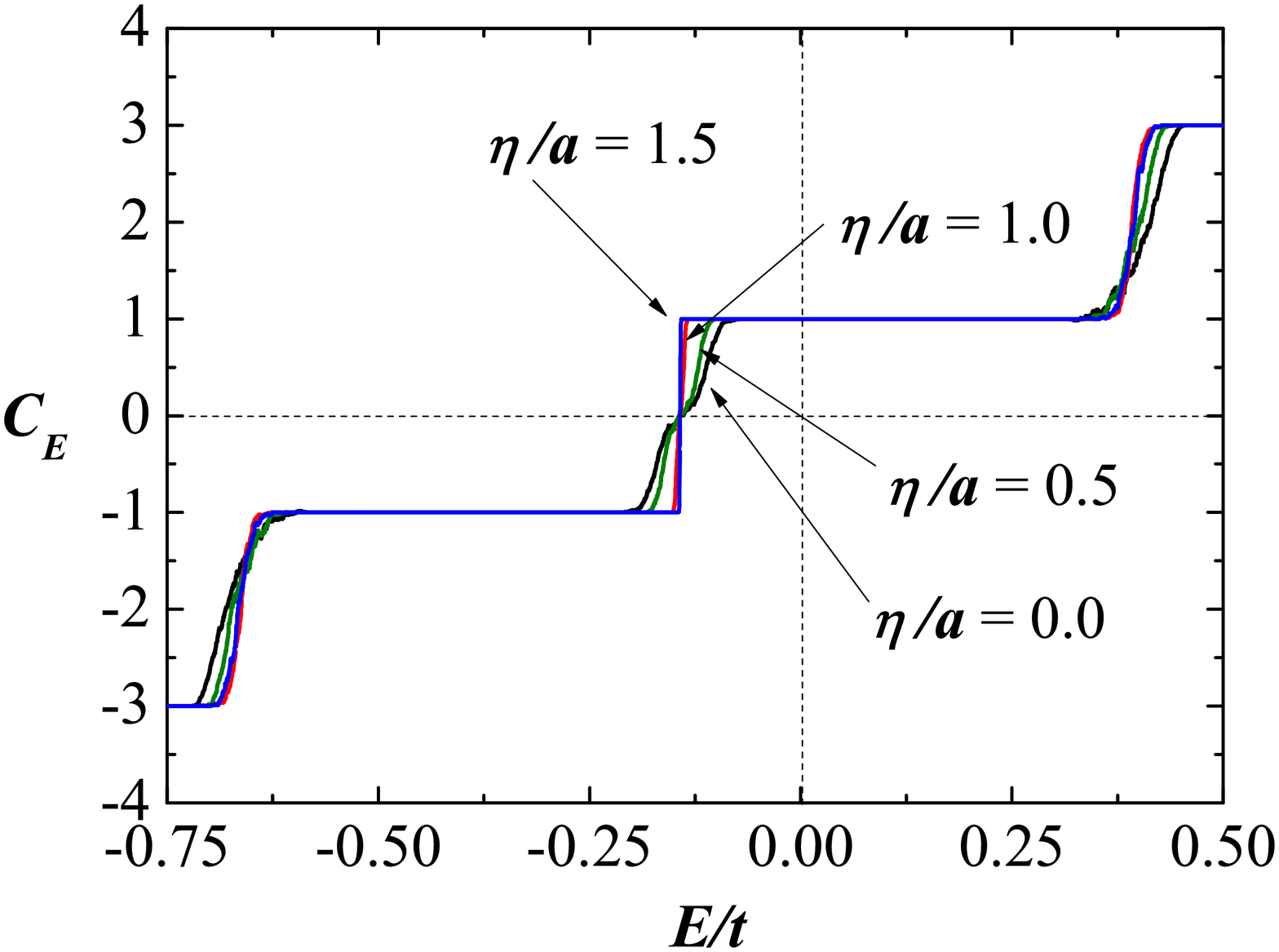}
\caption{(Color online) 
The Chern number $C_E$ (Hall conductivity in units of $e^2/h)$ as a function of the Fermi energy $E/t$ 
in the presence of the next-nearest-neighbor hopping $t'/t=0.05$ for $\eta/a=0.0,0.5,1.0$ and $1.5$, for 
$\phi/\phi_0 = 1/36$, $\sigma/t = 0.115$ and $(L_x/(\sqrt{3}a/2),L_y/(3a/2)) =(24,18)$.
\label{fig3}
}
\end{figure}

\begin{figure}
\includegraphics[scale=0.3]{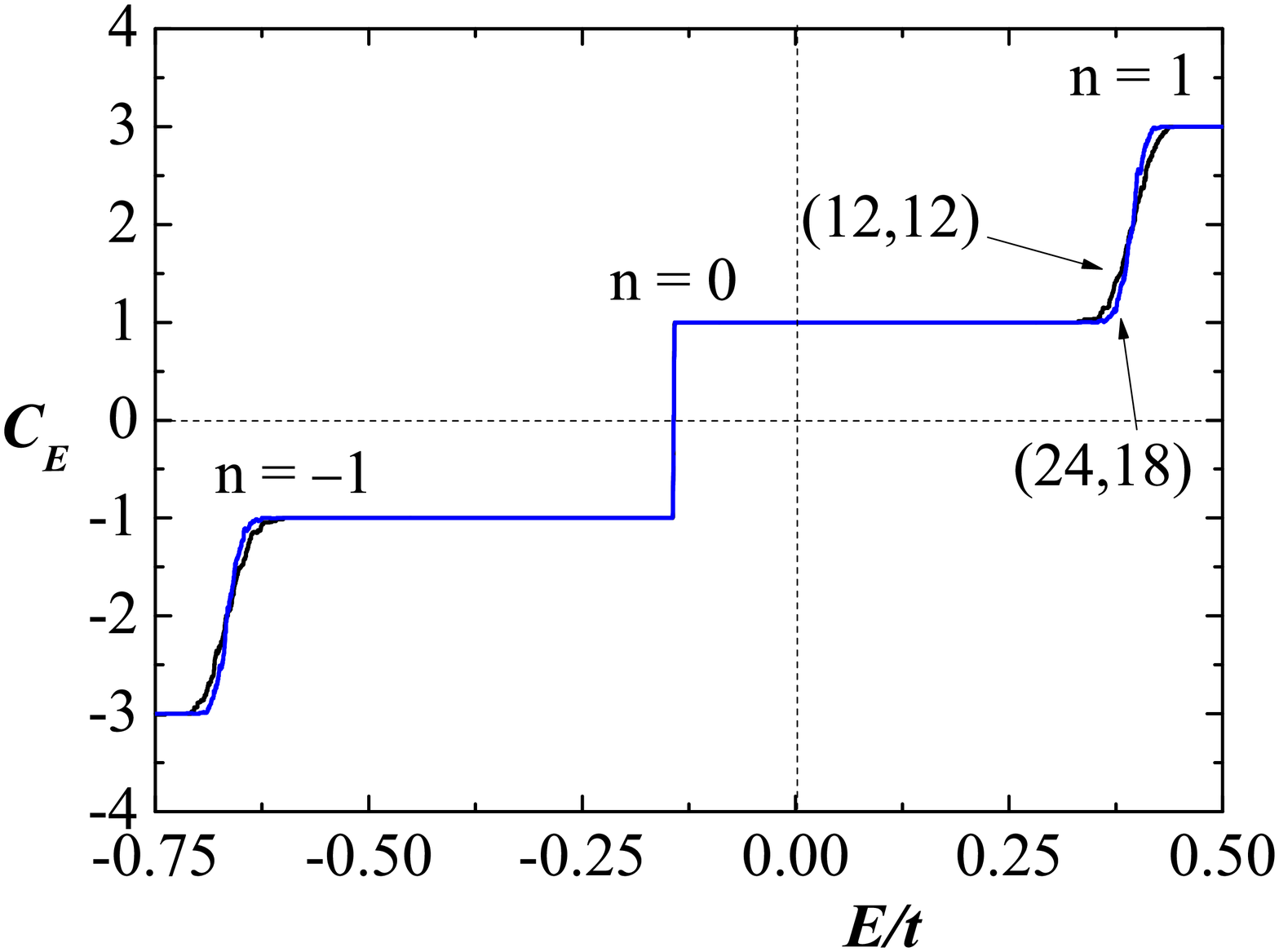}
\caption{(Color online) 
The Chern number $C_E$ (Hall conductivity in units of $e^2/h)$ as a function of the Fermi energy $E/t$ 
in the presence of the next-nearest-neighbor hopping $t'/t=0.05$ for $\eta/a=1.5$, for 
$\phi/\phi_0 = 1/36$, $\sigma/t = 0.115$ and $(L_x/(\sqrt{3}a/2),L_y/(3a/2)) =(24,18)$ and $(12,12)$. 
\label{fig4}
}
\end{figure}

\begin{figure}
\includegraphics[scale=0.3]{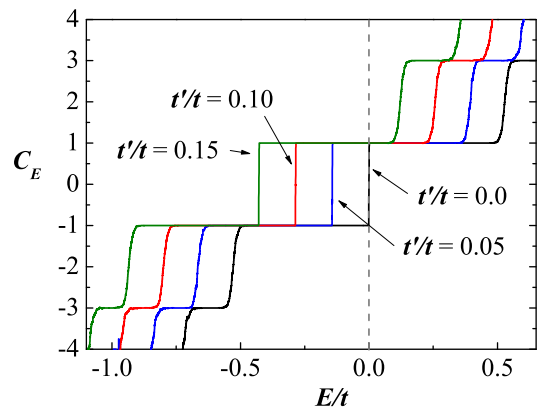}
\caption{(Color online) 
The Chern nmber $C_E$ (Hall conductivity in units of $e^2/h)$ as a function of the Fermi energy $E/t$ 
for various values of the next-nearest-neighbor hopping $t'$, with a 
fixed $\eta/a=1.5$, 
$\phi/\phi_0 = 1/36$, $\sigma/t = 0.115$ and $(L_x/(\sqrt{3}a/2),L_y/(3a/2)) =(24,18)$. 
\label{fig5}
}
\end{figure}

Next we show what happens when we switch on the next-nearest neighbor hopping $t' \neq 0$.  
Note that
the Dirac cones still exist at $K$ and $K'$ at an energy, $-3t'$, 
shifted from zero.\cite{Neto}  
The Hall conductivity in the case of  $t'/t = 0.05$ is shown in Fig. \ref{fig3} for several values of the 
disorder correlation length $\eta$.
It is clearly seen that the Hall plateau transition at the $n=0$ Landau level is again 
anomalously sensitive to the spatial correlation of the bond disorder. The results for different system sizes are 
displayed in  Fig.\ref{fig4}, where we see little size-dependence at $n=0$.  This occurs despite the fact that 
the chiral symmetry  of the lattice model is broken by the next-nearest neighbor 
hopping $t'\neq 0$. Consequently, the energy spectrum is not symmetric 
about $E=0$ with 
the $n=0$ Landau 
level shifted from $E=0$.    
If we vary the strength of the next-nearest neighbor hopping $t'$, the critical energy for the anomalous 
Hall transition associated with the $n=0$ Landau level is simply shifted 
with $t'$ (Fig.\ref{fig5}).  
For $n=\pm1$ Landau levels, 
by contrast, the usual scaling 
behavior (i.e., a narrower transition region for a larger system)  is 
observed (Fig. \ref{fig4}). 

The present results amount to that the anomalous transition remains at the energy of the Dirac cones, where 
the breakdown of the usual chiral symmetry does not 
affect the anomalous criticality although the energy 
spectrum is affected.  
This stability of the anomalous criticality 
is attributed to the preserved effective chiral symmetry 
for the $2 \times 2$ low-energy effective Hamiltonian in the vicinity of the Dirac cones.\cite{Neto}

\subsection{Tilted Dirac cones}

\begin{figure}
\includegraphics[scale=0.6]{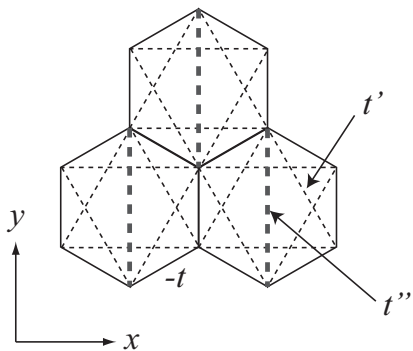}
\caption{The $\pi$ flux model having additional hopping 
$t''$ (thick dashed lines) in the $y$ direction across the hexagon of the honeycomb lattice. 
\label{fig6}
}
\end{figure}

To examine further the robustness of the anomaly at the $n=0$ Landau level, we consider a lattice model
having an third-neighbor hopping  $t''$ across the hexagon of the honeycomb lattice (Fig. \ref{fig6}). It has been shown \cite{HFA} that, 
for the case of $t'=0$,  
this model has also two Dirac cones in the first Brillouin zone for $-1<t"<3$, 
which are adiabatically connected to the Dirac cone 
in the original honeycomb lattice as 
the hopping $t"$ is continuously varied. When $t"$ is increased to 
$t$ (with $t'=0$), the model is equivalent to the $\pi$ flux model, which has been analyzed as a typical model with the Dirac dispersion.\cite{Hatsugai}
If we further introduce the second-neighbor hopping $t'\neq 0$ in 
the model, we have not only shift 
the energy of the Dirac cones, but also tilt the cones. 
For simplicity, here we confine ourselves to the case of $t"=t$, where 
two anisotropic Dirac cones are located at $(\pi/2\sqrt{3}a, \pi/2a)$ and $(-\pi/2\sqrt{3}a,5\pi/6a)$. In the presence of $t'$, 
Dirac cones are located at  $E=-2t'$.

The Hall conductivity obtained for this model is shown in Fig. \ref{fig8} in the presence of spatially correlated disorder with $\eta/a=1.5$.
For the case of $t'=0$, the system is chiral symmetric and two Dirac cones exist at $E=0$. We thus obtain 
anomalous criticality at $E=0$. When  the hopping $t'$ is introduced, 
thereby degrading the chiral symmetry, 
we clearly see that  the anomalous plateau transition 
are shifted in energy but preserved.  
The shift of the critical energy with $t'$ is again consistent with the energy shift of the Dirac cones.  
The result suggests that the anomaly in the $n=0$ Landau level is 
preserved even when the Dirac cones are tilted, 
although the tilt angle achieved in the present calculation is rather small (up to $\sim$ 5$^\circ$).

\begin{figure}
\includegraphics[scale=0.4]{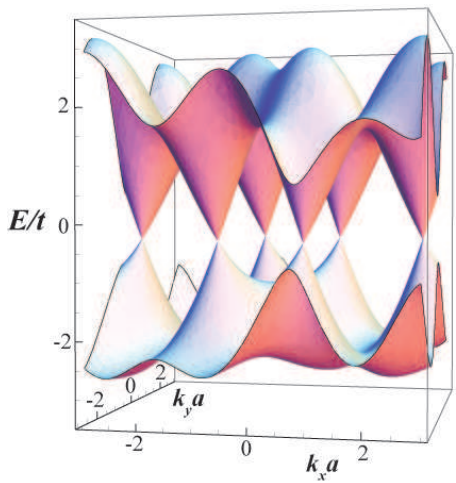}
\caption{(Color online) 
Tilted Dirac cones for the $\pi$-flux model with $t"=t$ and $t'/t=0.15$.  
The Dirac cones are located at 
$(k_x,k_y) =(\pi/2\sqrt{3}a, \pi/2a)$ and $(-\pi/2\sqrt{3}a,5\pi/6a)$ and 
$E=-0.3t$.
\label{fig7}
}
\end{figure}

\begin{figure}
\includegraphics[scale=0.3]{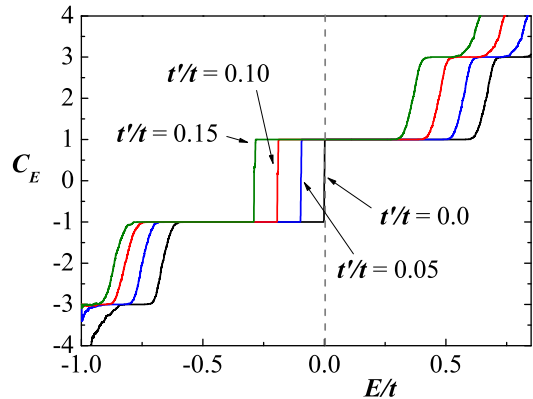}
\caption{(Color online) 
The Chern number $C_E$ (Hall conductivity in units of $e^2/h)$ as a function of the Fermi energy $E/t$ 
in the presence of the next nearest-neighbor hopping $t'$ for the $\pi$-flux model, for $\eta/a=1.5$, 
$\phi/\phi_0 = 1/36$, $\sigma/t = 0.115$ and $(L_x/(\sqrt{3}a/2),L_y/(3a/2)) =(18,12)$. The strength of the 
hopping $t"$ across the hexagon is assumed to be the average of $|t_{ij}|$  around the corresponding hexagon.  
\label{fig8}
}
\end{figure}

\section{AC Hall effect}
Now we would like to look at the ac extension of the Hall conductivity, 
i.e. optical Hall conductivity $\sigma_{xy}(\omega)$ 
to discuss the effect of the chiral symmetry.
The crucial interests is 
whether the existence or otherwise of the chiral symmetry 
dominates ac responses as well, where we pay particular 
attention to the plateau structure 
in the optical Hall conductivity $\sigma_{xy}(\omega)$.

So we have calculated $\sigma_{xy}(\omega)$,
where we solved the Hamiltonian containing various disorders
with the exact diagonalization method, since we are interested 
in the localization effects in the ac Hall conductivity.
We then  compute the optical Hall conductivity 
by using the Kubo formula, since the method 
for calculating the topological (integer)  Chern numbers 
in the dc case is not applicable to the optical conductivity.
For the ac Hall conductivity, we consider the usual honeycomb lattice with 
different types of disorder to examine how the symmetry of disorder 
is reflected in the ac conductivity.

The Kubo formula for the ac conductivity is given as
\begin{eqnarray}
\sigma_{xy}(\varepsilon_F,\omega)
&=&
\frac{i\hbar}{ L^2}
\sum_{\epsilon_a < \varepsilon_F}\sum_{\epsilon_b \ge \varepsilon_F}
\frac 1 {\epsilon_b-\epsilon_a}
\nonumber\\
&\times&
\left(\frac{j_x^{ab}j_y^{ba}}{\epsilon_b-\epsilon_a-\hbar\omega} 
-\frac{j_y^{ab}j_x^{ba}}{\epsilon_b-\epsilon_a+\hbar\omega}
\right) ,
\label{kuboformula}
\end{eqnarray}
where $\epsilon_a$ is the eigenenergy, $j_x^{ab}$ the current matrix 
elements between the eigenstates, 
and $\varepsilon_F$ the Fermi energy \cite{morimoto-opthall}.
The current operator is given by a gauge field derivative of the Hamiltonian as
$$
\mbox{\boldmath $j$}=\frac{\partial H}{\partial \mbox{\boldmath $A$}},
$$
where the gauge field and the Peierls phase in the Hamiltonians 
are related via 
$
2\pi\theta_{ij}=\mbox{\boldmath $A$}\cdot (\mbox{\boldmath $r$}_{i} -\mbox{\boldmath $r$}_{j})
$. 

To see the effect of the chiral symmetry on $\sigma_{xy}(\omega)$,
we have examined two cases: 
(i) The system with a site-potential disorder introduced 
as 
$$
H = \sum_{\langle i,j\rangle} -t e^{-2\pi{\rm i}\theta_{ij}} 
 c_i^{\dagger}c_j + {\rm h.c.} + \sum_{i} \delta\varepsilon_i c_i^\dagger c_i ,
$$
with a random potential $\delta\varepsilon$  having a Gaussian distribution. 
 The potential disorder, which breaks the chiral symmetry, 
models an effect of charged impurities in graphene samples.  
(ii) The system with random hopping, 
$$
H = \sum_{\langle i,j\rangle} (-t+\delta t_{ij}) e^{-2\pi{\rm i}\theta_{ij}} 
 c_i^{\dagger}c_j + {\rm h.c.} ,
$$
which preserves the chiral symmetry,  is considered to be coming from 
ripples in graphene samples.  
So we would look into the effect of these on the optical Hall conductivity.

\begin{figure}[tb]
\begin{center}
\begin{tabular}{c}
\includegraphics[width=0.9\linewidth]{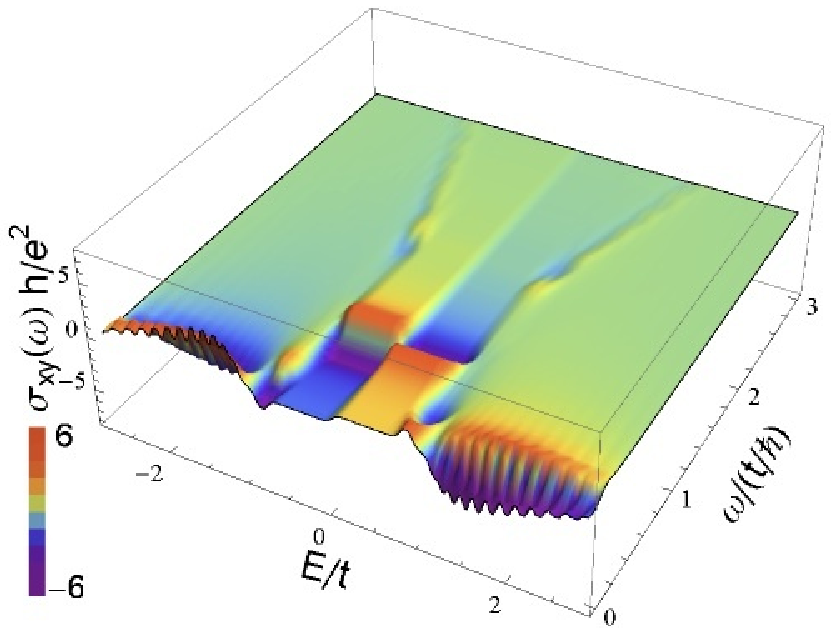}
\\
(a)\\
\includegraphics[width=0.8\linewidth]{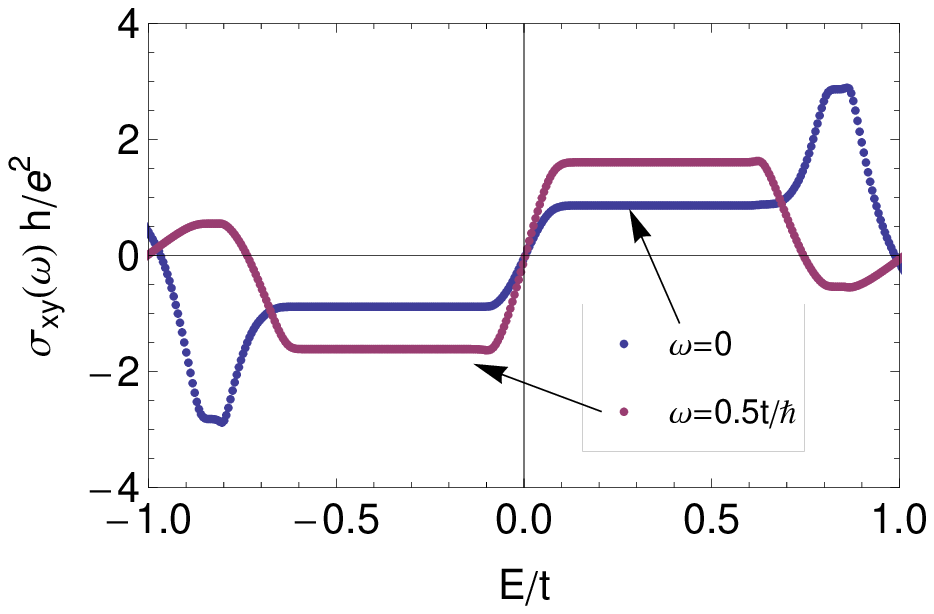}
\\
(b)\\
\end{tabular}

\end{center}

\caption{
(a) $\sigma_{xy}(\varepsilon_F, \omega)$ plotted against Fermi energy $\varepsilon_F$ 
and frequency $\omega$ for the honeycomb tight-binding model 
with a potential disorder $\sqrt{\langle \delta \varepsilon^2 \rangle}=0.1 t$ which breaks the chiral symmetry.  
(b) $\sigma_{xy}(\varepsilon_F, \omega)$ for $\omega=0,0.5 t/\hbar$ is displayed
to show the broadening of the step structure associated with 
the $n=0$ Landau level.  
The parameters are $\phi/\phi_0=1/18$, $(L_x/(\sqrt{3}a/2),L_y/(3a/2)) =(27,24)$.}
\label{ac-site}
\end{figure}

Fig.\ref{ac-site} shows $\sigma_{xy}(\varepsilon_F, \omega)$ plotted against Fermi energy $\varepsilon_F$ 
and frequency $\omega$ with a potential disorder 
$\sqrt{\langle \delta \varepsilon^2 \rangle}=0.1 t$ 
which breaks the chiral symmetry.  
The curve for $\omega=0$ in Fig.\ref{ac-site}(a) corresponds to dc Hall conductivity.
If we vary Fermi energy $\varepsilon_F$ for a fixed frequency $\omega$,
we notice that there are two distinct regions.  
For the Dirac-cone-like energy region bounded by 
the van Hove singularities ($|\varepsilon_F| < t$) 
we observe step structures at Dirac Landau levels, 
where we have wider plateaus for higher Landau levels 
due to a $\sqrt{B}$ dependence of cyclotron energy for Dirac QHE. 
At the van Hove singularities the Dirac QHE crosses over to the 
ordinary QHE for usual fermions around the band edges,
and outside the van Hove singularities ($|\varepsilon_F| > t$) an usual QHE step sequence is recognized.

If we look at the dependence on the frequency $\omega$, 
each plateau exhibits the cyclotron resonance behavior as 
$\omega$ is increased, where the resonance shape 
differs between the two regions.
For the usual QHE region the resonance occurs for small $\omega$ because of small cyclotron frequency ($\propto B$).  
In the Dirac QHE region, we have resonances 
for larger $\omega$ $(\propto \sqrt{B})$, which appear at various 
resonance frequencies that correspond to 
resonances between different for LL's with non-uniform LL spacings 
with a peculiar selection rule, $|n| \leftrightarrow |n+1|$ for Dirac QHE.  
An interesting point observed for the 
honeycomb tight-binding model is that these two regions emerge from the 
nature of the honeycomb band dispersion and separated by van Hove singularities.  

Fig.\ref{ac-site}(b) depicts the step structure around $n=0$ LL for dc and ac ($\omega=0.5 t/\hbar$) Hall conductivities, 
which shows that the  $n=0$ steps are blurred both for dc and ac 
response 
when the disorder does not respect  the chiral symmetry.

\begin{figure}[tb]
\begin{center}
\begin{tabular}{c}
\includegraphics[width=0.9\linewidth]{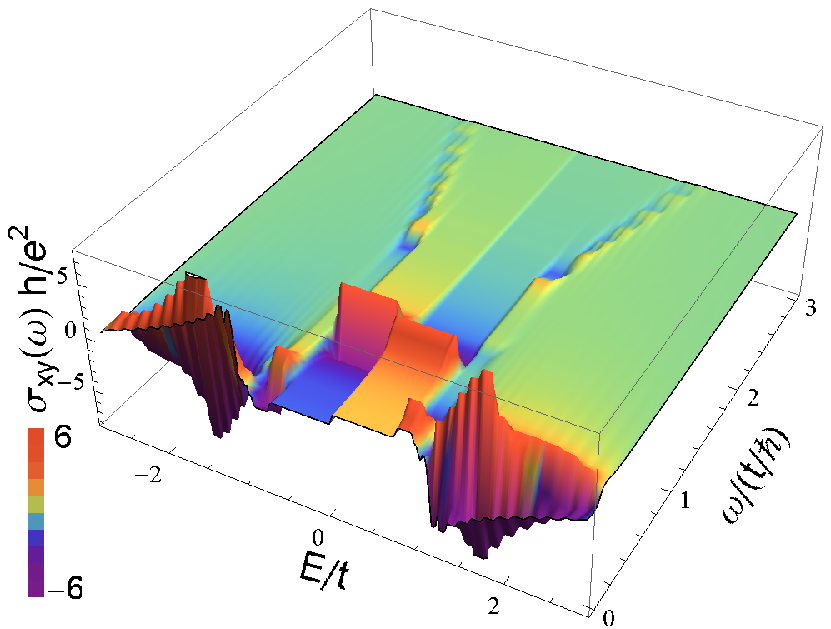}
\\
(a)\\

\includegraphics[width=0.8\linewidth]{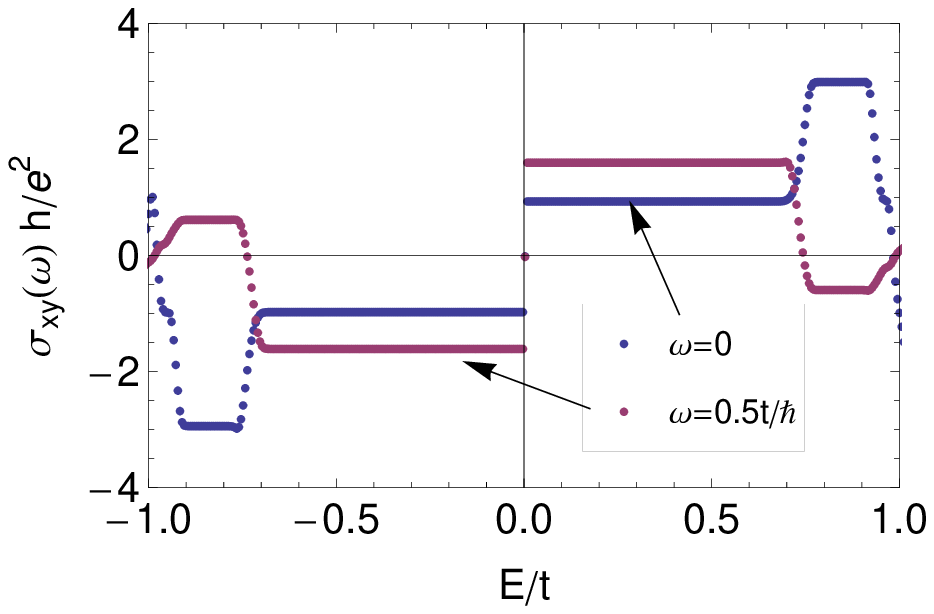}
\\
(b)\\
\end{tabular}

\end{center}

\caption{
(a) $\sigma_{xy}(\varepsilon_F, \omega)$ plotted against Fermi energy $\varepsilon_F$ 
and frequency $\omega$ for the honeycomb tight-binding model 
with random hopping $\sqrt{\langle \delta t^2 \rangle}=0.1 t$ which respects the chiral symmetry.
(b) $\sigma_{xy}(\varepsilon_F, \omega)$ with $\omega=0,0.5 t/\hbar$ is displayed
to show the anomalous step structure at n=0 LL retained for ac Hall conductivity.
The parameters involved are $\eta/a=2.0$, $\phi/\phi_0=1/18$, $(L_x/(\sqrt{3}a/2),L_y/(3a/2)) =(27,24)$.
}
\label{ac-bond}
\end{figure}

Now we come to the original question of what happens to the 
ac Hall conductivity when the disorder respects the chiral 
symmetry.  Figure \ref{ac-bond} shows $\sigma_{xy}(\varepsilon_F, \omega)$ 
for the case of a random hopping $\sqrt{\langle \delta t^2 \rangle}=0.1 t$ which respects the chiral symmetry.  
The overall structure of the $\sigma_{xy}(\varepsilon_F, \omega)$ is similar to the random potential case, but we can immediately 
notice that  the $n=0$ step in $\sigma_{xy}(\varepsilon_F, \omega)$ 
exhibits an anomalously sharp step structure at even for the 
optical Hall conductivity as in the dc Hall conductivity.
Indeed,  Fig.\ref{ac-bond}(a) indicates that 
$\sigma_{xy}(\varepsilon_F=0, \omega)$ shows a sharp step for all 
the values of $\omega$, which should come from 
the protected $n=0$ LL by the chiral symmetry.

We can conclude that the chiral symmetry and the zero mode protection give a crucial effect on the  ac response of the system as in the dc response,
where  the presence of the chiral symmetry gives a rise to the anomalously sharp step structure in $\sigma_{xy}(\varepsilon_F, \omega)$ associated with delta-function like $n=0$ LL.  
The honeycomb lattice calculation, performed here for the optical Hall conductivity,  is not only more realistic than the effective Dirac field description, 
but also reveals an existence of  
two regions (Dirac QHE and usual fermionic QHE) coming from the 
van Hove singularity in the band dispersion, where totally 
different plateau width and cyclotron resonance structures are 
observed.

\section{Conclusions}

We have investigated, based on several lattice models, the relationship between the chiral symmetry and 
the anomalous criticality associated with the $\delta$-function like density of states at the $n=0$ Landau level in graphene, which yields a 
step-function like Hall plateau transition. The ac as well as dc Hall conductivity has been 
examined as a function of the Fermi energy in the presence of  the spatially-correlated chiral symmetric bond disorder.
For the dc Hall conductivity, it has been shown that the anomalous criticality
is surprisingly robust against the introduction of the next-nearest neighbor hopping in the honeycomb lattice, which is considered to be 
appreciable in real graphene. The anomalous criticality may therefore be realized 
in a suspended clean graphene where ripples are the major disorder.
As for the symmetry, the present results implies that the anomalous criticality can be insensitive to the breakdown of 
the usual chiral symmetry 
as long as the effective chiral symmetry that produces the massless 
Dirac cones is preserved.  
We have also shown that the anomaly is robust against a slight tilt in the Dirac cones, which suggests the possibility to 
observe the anomalous criticality not only in graphene but also in organic materials.
For the ac Hall conductivity, we have shown that even for finite frequencies, the transition at the $n=0$ Landau level 
exhibits the anomalous criticality, with the robust plateau structure 
retained  for long-ranged, chiral-symmetric bond disorders.  
Based on these numerical results, we conclude that the chiral symmetry manifests itself both in the dc and the ac Hall transitions yielding the 
anomalous criticality at the $n=0$ Landau level of graphene.  

\begin{acknowledgments}
We wish to thank Yoshiyuki Ono, Tomi Ohtsuki 
for useful discussions and comments.
The work was supported in part by grants-in-aid for scientific research,
Nos. 20340098 (YH and HA) and 22540336 (TK)
 from JSPS and
No. 22014002 (YH) 
on priority areas from MEXT. 

\end{acknowledgments}


\vfill
\end{document}